\documentstyle[12pt,epsf]{article}
\newcommand{\del}{{\partial}}
\newcommand{\beq}{\begin{eqnarray}}
\newcommand{\eeq}{\end{eqnarray}}
\newcommand{\be}{\begin{eqnarray*}}
\newcommand{\ee}{\end{eqnarray*}}
\newcommand{\bk}{{\bf k}}

\newcommand{\br}{{\bf r}}

\newcommand{\ra}{\rightarrow}

\newcommand{\ve}{\varepsilon}

\newcommand{\ex}[1]{\langle\,#1\,\rangle}

\newcommand{\f}[2]{\mbox{$\frac{\scriptstyle#1}{\scriptstyle#2}$}}
\newcommand{\2}{\f{1}{2}}

\begin{document}

\centerline{\Large\bf Ideal quantum gases in two dimensions}
\vskip 10mm
\centerline{S. Viefers, F. Ravndal and T. Haugset}
\centerline{\it Institute of Physics}
\centerline{\it University of Oslo}
\centerline{\it N-0316 Oslo, Norway}

\vspace{10mm}
\centerline{To appear in Am. J. Phys.}

\vspace{10mm}
{\bf Abstract:} {\footnotesize Thermodynamic properties of non-relativistic
bosons and fermions in two spatial dimensions and without interactions are
derived. All the virial coefficients are the same except for the second, for
 which the signs are opposite. This results in the same specific heat for the
two gases. Existing equations of state for the free anyon gas are also
discussed and shown to break down at low temperatures or high densities.}

\vspace{10mm}
\section{Introduction}
Physics of two-dimensional systems used to be of primarily academic interest
in
providing mathematically simpler versions of realistic problems
in three dimensions. The fundamental discovery of Leinaas and Myrheim in
1977 \cite{LM} of the possibility for intermediate quantum statistics for
identical
particles in two dimensions interpolating between standard Bose-Einstein and
Fermi-Dirac statistics therefore generated initially little general interest.
Particles with this new statistics were later named anyons by Wilczek
\cite{FW1}.
He showed that they could be modelled as carrying both a charge and a
magnetic flux locked together in a very special way.

With the progress of modern microelectronics this situation changed
\cite{Challis}.
The discovery of the integer quantum Hall effect \cite{QHE} could be
explained by the
special quantum effects of electrons effectively confined between layers
of different materials in planar transistors. Later it became
clear that excitations in the fractional quantum Hall effect obeyed
intermediate statistics \cite{FQHE}. This in turn suggested that the
mechanism behind the new, high-temperature superconductivity seen in layers of
copperoxide planes in different materials, could also be anyonic
\cite{RBL,SC,FW2}.

Even if anyons should not turn out to be of great practical importance in
the coming years, two-dimensional systems in general will play an important
role in microelectronics and material sciences. Since the third dimension
is effectively frozen out only at very low temperatures, it is also obvious
that quantum effects will govern the behaviour of these new systems.

Here we consider the thermodynamics of ideal quantum gases
in two dimensions. Ordinary bosons and fermions are treated in the next
section where their equation of state is derived. In Section 3 we review
the basic properties of anyons and solve the simplest case of two anyons
bound by a harmonic potential. The resulting energy spectrum is then used
in Section 4 to calculate the second virial coefficient for anyons.
Including some recent results for the higher virial coefficients, we then
discuss in the last section their full equation of state.

\vspace{10mm}
\section{Statistical mechanics of bosons and fermions}

We consider $N$ free and identical particles of mass $m$ in a box of
two-dimensional volume $V$.
Each particle is characterized by the momentum $\bk$ and corresponding energy
$\ve = \bk^2/2m$. When they are in thermal equilibrium at temperature
$T$ and chemical potential $\mu$, their thermodynamics follows from the
grand canonical partition function
\beq
    \Xi = \prod_\bk [1 \pm e^{-\beta(\ve_\bk - \mu)}]^{\pm 1},      \label{2.1}
\eeq
where the upper signs are for Fermi-Dirac statistics and the lower ones for
Bose-Einstein statistics \cite{Huang}. Here $\beta = 1/k_B T$ where $k_B$ is
the Boltzmann constant. In this ensemble the corresponding free energy density
is simply the pressure $P = \ln{\Xi}/\beta V$ which becomes
\beq
    \beta P = \pm {1\over V} \sum_\bk
    \ln{[1 \pm e^{-\beta(\ve_\bk - \mu)}]}.                       \label{2.2}
\eeq
The particle density $\rho = N/V$ then follows from
\beq
    \rho = \left({\del P\over \del\mu}\right)_\beta
         = {1\over V} \sum_\bk {1\over e^{\beta(\ve_\bk - \mu)} \pm 1} .
                                                                  \label{2.3}
\eeq
Similarly, if $E$ is the total energy of the particles, the energy density
${\cal E} = E/V$ is given by the derivative
\beq
    {\cal E} =	-\left({\del \over \del\beta}\,\beta P\right)_{\beta\mu}
             =	 {1\over V} \sum_\bk {\ve_\bk\over e^{\beta(\ve_\bk - \mu)}
 \pm 1} .
          \label{2.4}
\eeq
Since the lowest one-particle energy is here $\ve = 0$, the chemical
 potential can
not become negative for bosons. Their fugacity $z = \exp{(\beta\mu)}$ will
thus
always be less than one. No such upper limit exits for fermions.

In two dimensions we can replace the momentum sum by the integral
\beq
    \sum_\bk \ra V\!\int\!{d^2k\over (2\pi\hbar)^2}                \label{2.5}
\eeq
in the thermodynamic limit where $V \ra \infty$. Writing $kdk = md\ve$,
we see that the particle density (\ref{2.3}) is given by the integral
\beq
	 \rho = {m\over 2\pi\hbar^2}\int_0^\infty \!d\ve
           {1\over e^{\beta(\ve - \mu)} \pm 1}
		  = \pm {m\over 2\pi\beta\hbar^2}\log{(1 \pm z)}.  \label{2.6}
\eeq
Introducing the thermal wavelength $\Lambda = (2\pi\beta\hbar^2/m)^{1/2}$,
we can write the result for the fugacity as
\beq
    z = \mp\left(1 - e^{\pm\rho\Lambda^2}\right).                  \label{2.7}
\eeq
The upper signs are still for fermions and the lower ones for bosons.
For bosons $z$ smoothly approaches the value one as the temperature goes to
zero. Thus there is no Bose-Einstein condensation in two dimensions except
at $T = 0$ when all the particles are in the ground state.

For the energy density (\ref{2.4}) we similarly find in the thermodynamic limit
\beq
    {\cal E} = {\beta\over\Lambda^2}\int_0^\infty \!d\ve
               {\ve\over e^{\beta(\ve - \mu)} \pm 1}  .
  \label{2.8a}
\eeq
A partial integration where the boundary term vanishes then gives
\beq
    {\cal E} = \pm {1\over\Lambda^2}\int_0^\infty \!d\ve
               \ln{[1 \pm  e^{-\beta(\ve - \mu)}]}    .
  \label{2.8b}
\eeq
This is seen to equal the pressure (\ref{2.2}) in the same limit. We thus have
$P = {\cal E}$. In three dimensions the corresponding result is the more
well-known relation $P = (2/3){\cal E}$ for non-relativistic particles.

The integral (\ref{2.8b}) for the pressure can be expressed in a more compact
 way.
Changing variable of integration to $t = 1 \pm z\exp{(-\beta\ve)}$ we obtain
\beq
    \beta P(\beta,\mu) = \mp{1\over\Lambda^2}\,\mbox{Li}_2(1 \pm z)
 \label{2.7a}
\eeq
where $\mbox{Li}_2(x)$ is the dilogarithmic function defined by \cite{AS}
\beq
     \mbox{Li}_2(x) =	- \int_1^x\!dt\,{\log{t}\over t - 1}.
   \label{2.7b}
\eeq
When $0 < x \le 2$ one can obtain a convergent series for the pressure by
expanding the logarithm in (\ref{2.8b}) and integrating term by term. This
expansion
is thus valid at all temperatures for bosons and only at high temperatures for
fermions.

With the explicit solution (\ref{2.7}) for the fugacity we can now obtain a
differential
equation for the pressure which will give the equation of state. Forming
the partial derivative
\[
   \left({\del P\over\del\rho}\right)_T =
   \left({\del P\over\del\mu}\right)_T\left({\del\mu\over\del\rho}\right)_T,
\]
we see that the first derivative on the right-hand side is just the density
(\ref{2.3}) and the second is obtained from (\ref{2.7}) as
\[
    \left({\del\mu\over\del\rho}\right)_T = \mp{\Lambda^2\over\beta}
    {1\over e^{\mp\rho\Lambda^2} - 1}.
\]
The equation of state is thus explicitly given by the integral
\beq
    \beta P = \mp \int_0^\rho\!d\rho'
    {\rho'\Lambda^2 \over e^{\mp\rho'\Lambda^2} - 1},          \label{2.10}
\eeq
as first obtained by Sen \cite{DS}.

At high temperatures we can express the equation of state in terms of the
virial
coefficients. They can now be obtained by using
the expansion
\beq
      {x\over e^x - 1} = \sum_{n=0}^\infty B_n\,{x^n\over n!}    \label{2.11}
\eeq
which is convergent for $0 < |x| < 2\pi $ \cite{AS}. It defines the Bernoulli
numbers $B_n$,
giving $B_0 = 1$, $B_1 = -1/2$, $B_2 = 1/6$, $B_3 = 0$, $B_4 = -1/30$ etc.
with
$B_{2n+1} = 0$. We then obtain from (\ref{2.10}) the virial expansion
\beq
      \beta P = \sum_{\ell = 1}^\infty A_\ell\,\rho^\ell \Lambda^{2(\ell - 1)}
                                                                \label{2.12}
\eeq
where the dimensionsless virial coefficients are simply given in terms of the
Bernoulli numbers \cite{DS,SV},
\beq
      A_\ell = (\mp)^{\ell - 1} {B_{\ell - 1}\over \ell !}.      \label{2.13}
\eeq
The lowest ones are $A_2 = \pm1/4$, $A_3 = 1/36$, $A_4 = 0$, etc. and give
for the equation of state at high temperatures or low densities
\beq
     \beta P = \rho \pm {1\over 4}\rho^2\Lambda^2 + {1\over 36}\rho^3\Lambda^4
      - {1\over 3600}\rho^5\Lambda^8 + {1\over 211680}\rho^7\Lambda^{12}
      + \cdots   .                                             \label{2.14}
\eeq
Notice the interesting fact that since all the odd Bernoulli numbers are zero
after $B_1$, all the even virial coefficients are zero after $A_2$. Hence,
the only difference between the pressures in the bosonic or fermionic gases
 comes
from the second term in the expansion (\ref{2.14}), i.e.
\beq
      P_{F} - P_{B} = {1\over 2\beta}\rho^2\Lambda^2
                      = {\hbar^2\over m}\pi\rho^2  .           \label{2.15}
\eeq
The result does not depend on the convergence
of the virial expansion. It follows directly from (\ref{2.7a}) when we make
use of the special property of the dilogarithmic function\cite{AS}
\beq
     Li_2(x) + Li_2(x^{-1}) = -\2\log^2{x},                  \label{2.16}
\eeq
where $x = \exp{(\rho\Lambda^2)}$ in our case.	The difference is just the
pressure or energy density ${\cal E}_0 = \hbar^2\pi\rho^2/m$ in the Fermi
gas at zero temperature as shown in the Appendix. Since it is constant with
respect to $T$,
we see that the specific heats of these ideal quantum gases are equal as
pointed out by Aldrovandi \cite{RA}. It would be of interest to have a more
 direct
understanding of this simple result.

We have plotted the pressures of the gases as a function of temperature at
fixed density in Fig.1. At high temperatures we can use the
virial expansion (\ref{2.12}). Since it only converges for $\rho\Lambda^2
< 2\pi $, it breaks down at low temperatures. For the Fermi
gas just above $T = 0$ we can instead use the Sommerfeld
expansion \cite{Huang} derived in the Appendix. In two dimensions it is found
 to
terminate after the second order. At slightly higher temperatures it
smoothly matches up with the virial expansion. We can obtain the pressure
of the Bose gas by just subtracting the constant term ${\cal E}_0$ from
the Fermi pressure. In the Appendix we have also derived exact equations of
state
for these gases as explicit functions of the dimensionless quantity
$\rho\Lambda^2$
combining the results in (\ref{2.7}) and (\ref{2.7a}).

\section{Quantum mechanics of anyons}

The wavefunctions of bosons and fermions differ under the interchange of two
or more particles. While the wavefunction for bosons is completely symmetric,
the wavefunction for fermions is completely antisymmetric. An interchange of
two particles described by the wavefunction $\psi(\br_1,\br_2)$ is shown in
Fig.2a. Their relative vector is rotated by the angle
$\Delta\phi = \pi$. The wavefunction is then transformed into
$\psi(\br_2,\br_1)$. Since the two particles are indistinguishable, this
wavefunction can only differ by a phase angle $\theta$ from the initial
wavefunction,
\beq
    \psi(\br_2,\br_1) = e^{i\theta}\psi(\br_1,\br_2).          \label{3.1}
\eeq
Had the interchange taken place in the opposite direction as in Fig.2b, the
final wavefunction would have been
\beq
    \psi'(\br_2,\br_1) = e^{-i\theta}\psi(\br_1,\br_2)         \label{3.2}
\eeq
since these two exchanges followed by each other are equivalent to no exchange.

In three dimensions we now see that this latter interchange is equivalent to
the first since it can be obtained by an out-of-page rotation around the
line connecting the two particles in their final position. Thus we have
$\psi'(\br_2,\br_1) =\psi(\br_2,\br_1)$ which implies that
$\exp({2i\theta}) = 1$. This equation has the two solutions
$\theta = 0~mod~2\pi$ and $\theta = \pi~mod~2\pi$. In the first case the
wavefunction is symmetric under the interchange and the particles are bosons,
while in the other case it is antisymmetric and the particles are fermions.

In two dimensions the two interchanges cannot be related for particles and the
parameter $\theta$ can have any value in the interval $0 \le \theta < 2\pi$.
We then have the possibility for intermediate statistics which interpolates
 between
Bose-Einstein and Fermi-Dirac statistics. The wavefunction for the
corresponding particles which are called anyons, will thus be neither
completely
symmetric nor antisymmetric. Needless to say, the statistical
mechanics of anyon gases will be much different from the behaviour of
bosons and fermions \cite{Anyons,AL}.

A quantum mechanical description of $N$ anyons can be obtained by modifying
the classical Lagrangian
\beq
    L = {1\over 2}\sum_{i=1}^N m\dot{\br}_i^2 - V(\br_1,\br_2,\cdots,\br_N)
                                                                   \label{3.3}
\eeq
where the last term describes the potential energy of the particles and is
symmetric in all its arguments. The quantum mechanical motion of this
system is now given by the integration of $\exp{(i\int dt L/\hbar)}$ over all
possible
paths the system can take \cite{RPF}. When the motion corresponds just to an
interchange of two particles, the net effect should be a change of the
wavefunction by a phase as in (\ref{3.1}). This effect can now be built into
the above Lagrangian by adding a topological term
\beq
     L \ra L_\theta = L + {\theta\over\pi}\hbar\sum_{i<j}\dot{\phi}_{ij}.
                                                                  \label{3.4}
\eeq
This additional term is the sum of time derivatives of the relative angles of
all pairs of particles in the system as illustrated in Fig.3. These angles
obviously depend on the positions of the particles. When the angle $\phi_{ij}$
changes by $\Delta\phi_{ij} = \pi$, the corresponding change in the
wavefunction
is
\[
     e^{i{\theta\over\pi}\int_0^\pi d\phi} = e^{i\theta},
\]
as desired. From this modified Lagrangian we can obtain the canonical
momenta and quantization can be performed by standard methods \cite{AL}.

In this way we can describe anyons as ordinary particles in two dimensions
with
a special, topological interaction. Bosons obey symmetric statistics and
prefer to be in the same quantum state. Fermions can be considered as bosons
with a statistical interaction  of strength $\alpha\equiv\theta/\pi = 1$.
This results in an effective repulsion between fermions which is usually
explained by the Pauli principle. Similarly, one can describe bosons as
fermions with an additional, statistical interaction with $\alpha = 1$. The
effect of this topological interaction for arbitrary values of the
parameter $\alpha$ in systems of many anyons is not yet known.

Essentially the only known system where exact results can be derived is
the case of two anyons in a harmonic oscillator potential
\beq
    V(\br_1,\br_2) = \2 m\omega^2(\br_1^2 + \br_2^2).         \label{3.5}
\eeq
This is equivalent to the case of two anyons in a uniform magnetic field, see
e.g. \cite{Johnson}.
Introducing the center-of-mass coordinate ${\bf R} = (\br_1 + \br_2)$ and
the relative coordinate $\br = \br_1  - \br_2 = (r\cos{\phi},r\sin{\phi})$,
the Lagrangian (\ref{3.4}) becomes
\beq
    L_\theta = \2 M \dot{\bf R}^2 + \2 M\omega^2\,{\bf R}^2
 			  + \2 \mu({\dot r}^2 + r^2{\dot\phi}^2 +
\omega^2 r^2)
             + \alpha\hbar{\dot\phi}.                          \label{3.6}
\eeq
We have here introduced the total mass $M = 2m$ and the reduced mass
$\mu = m/2$. The statistical parameter enters only in the relative motion
described
by the last term in the
Lagrangian. The center-of-mass motion is described by the first part which
is just an ordinary, two-dimensional harmonic oscillator. In order to
quantize the system we need the canonical momentum $p_r = m{\dot r}$
and
\beq
     p_\phi = {\del L\over\del{\dot\phi}} = \mu r^2{\dot\phi} + \alpha\hbar.
                                                               \label{3.7}
\eeq
They allow the construction of the Hamiltonian function of the system. In the
quantum
description the relative motion is governed by the operator
\beq
     H = - {\hbar^2\over 2\mu}
           \left[{1\over r}{\del\over\del r}r{\del\over\del r}
       + {1\over r^2}\left({\del\over\del\phi} - i \alpha\right)^2\right]
       + \2\mu\omega^2\,r^2.                                    \label{3.8}
\eeq
The Schr\"odinger equation is separable, and the corresponding wave
functions can be written as
\beq
     \psi(r,\phi) = e^{im\phi}R(r),                            \label{3.9}
\eeq
where the angular quantum number $m$ is an even number if the particles with
the
statistical interaction are bosons and an odd number if they are fermions.
For a given angular momentum the radial Schr\"odinger equation then
simplifies to
\beq
      \left(-{\hbar^2\over 2\mu}
      \left[{d^2\over dr^2} + {1\over r}{d\over dr}
      + {1\over r^2}(m - \alpha)^2\right]  + \2\mu\omega^2\,r^2\right)R(r)
      = E R(r)                                                 \label{3.10}
\eeq
which has exactly the same form as for an ordinary, two-dimensional harmonic
oscillator with a angular momentum $\ell = m - \alpha$. The eigenvalues
are thus \cite{LM}
\beq
       E_{nm} = \hbar\omega[2n + |m - \alpha| + 1],            \label{3.11}
\eeq
where $n = 0,1,2,\cdots$ is the radial quantum number. If the particles are
bosons,
the angular momentum takes the values $m = 0, \pm 2, \pm 4, \cdots$. The
quantized energy
levels as functions of the statistical parameter are shown in Fig.4. They are
seen to
repeat themselves for other values of $\alpha$ with period 2 although the
energy
of each indiviual state with definite quantum numbers $(n,m)$ is not periodic.

{}From Fig.4 we see that the energy levels fall into two separate classes
with opposite slopes. Within each class the degeneracy increases linearly
with the number of the level counted from below. For more than two
anyons we only know the lowest energy levels from numerical calculations
\cite{Sporre}. Again one finds that they can be grouped into different
classes depending on their slopes as functions of $\alpha$. This can be
explained by crude, semi-classical considerations \cite{Ruud}. In the case
of $N$ anyons, exact expressions for the ground state and some excited
states have been found, see for example \cite{Johnson}.
Apart from that, few exact results are known for systems of more than two
anyons.

\vspace{10mm}
\section{The second virial coefficient}

Gases of fermions or bosons at very high temperatures behave as classical,
ideal
gases, independent of the quantum statistics. We expect to see the same
behaviour in a gas of anyons. At lower temperatures quantum effects come
into play depending on the statistics. They result in a non-zero value
for the second virial coefficient, which for a gas confined to a volume $V$,
can be calculated from \cite{Huang}
\beq
    B_2 = \lim_{V\ra\infty}{V\over 2}\!\left(1 - 2\,{Z_2\over Z_1^2}\right)
 \label{4.1}
\eeq
where $Z_1$ and and $Z_2$ are the partition functions of one and two particles
respectively. While $Z_1$ is independent of statistics and thus the same
for all free particles, $Z_2$ for anyons is more difficult to derive.
It was first calculated by Arovas et al. \cite{Arovas} using Feynman's path
integral formalism \cite{RPF}. Since we know all the energy levels
(\ref{3.11})
of the two-anyon system, it can also be directly obtained from \cite{Comtet}
\beq
    Z_2 = Z_1\sum_{nm}e^{-\beta E_{nm}}
        = Z_1\sum_{n=0}^\infty e^{-\beta\hbar\omega(2n+1)}\sum_{s=-\infty}^
\infty
          e^{-\beta\hbar\omega|2s - \alpha|}                     \label{4.2}
\eeq
where we have written the angular momentum $m=2s$ when the anyons are
described in terms of bosons. The factor $Z_1$ in front gives the
contribution from
the center-of-mass motion. From (\ref{3.6}) we see that it equals the
partition function for one particle in a two-dimensional harmonic oscillator
potential, i.e.
\beq
    Z_1 = \left(\sum_{n=0}^\infty e^{-\beta\hbar\omega(n+\2)}\right)^2
        = {1\over 4\sinh^2(\beta\hbar\omega/2)}.                    \label{4.3}
\eeq
In (\ref{4.2}) the double sum is just the product of two such geometric series
and gives similarly
\beq
    Z_2 = Z_1 {\cosh\beta\hbar\omega(1-\alpha)\over 2\sinh^2(\beta\hbar\omega)}
					                          \label{4.4}
\eeq
for two anyons in a harmonic oscillator potential.

The infinite volume limit $V\ra \infty$ in (\ref{4.1}) is replaced by the
zero slope limit $\omega\ra 0$ for particles confined instead by a harmonic
potential. In this limit (\ref{4.3}) approaches
$Z_1 \ra 1/(\beta\hbar\omega)^2$
which should be compared with the result for a free particle in a box of volume
$V$,
\beq
    Z_1 = V\int\!{d^2 k\over(2\pi\hbar)^2}e^{-\beta\bk^2/2m}
        = {V\over\Lambda^2}                                       \label{4.5}
\eeq
where $\Lambda$ again is the thermal wavelength. Comparing these two results,
we have the relation
\beq
    V = \left({\Lambda\over\beta\hbar\omega}\right)^2              \label{4.6}
\eeq
In addition, in the standard formula (\ref{4.1}) we must make the replacement
$V/2 \ra V$ when the particles
are confined by an oscillator potential as here \cite{Kaare}. Expanding the
last
term to second order in $\omega$, we obtain
\be
   2\,{Z_2\over Z_1^2}
   & = &  \left[1 + \2(\beta\hbar\omega)^2(1 - \alpha)^2 + \cdots\right]
          \left[1 - \f{1}{4}(\beta\hbar\omega)^2 + \cdots\right]  \\
   & = & 1 - \left[\f{1}{4} - \2(1 - \alpha)^2\right](\beta\hbar\omega)^2 +
\cdots
\ee
The second virial coefficient for anyons can thus be written as $B_2 =
A_2\Lambda^2$ where its dimensionless value is
\beq
    A_2 = \f{1}{4} - \2(1 - \alpha)^2.                            \label{4.7}
\eeq
For ordinary bosons which have $\alpha = 0$, we recover the result $A_2 = -1/4$
from Section 2, while $\alpha = 1$ gives the standard fermion result
$A_2 = 1/4$. $A_2$ is plotted as a function of the statistical parameter in
Fig.5a where it is seen to interpolate smoothly between the fermionic and
bosonic values. The cusps at the bosonic points are found to disappear when
there is an additional interaction potential between the anyons depending on
the separation as $1/r^2$ \cite{Cusp}.

\vspace{10mm}
\section{Equation of state for anyons}
In order to calculate the higher virial coefficients $B_3$, $B_4,\ldots$, we
need the
partition functions for $N=3, 4,\ldots$ anyons. But the energy levels and the
corresponding degeneracies for these systems are not known. Instead of such
exact
calculations one has to rely on perturbation theory or numerical methods.
Following the latter approach, Myrheim and Olaussen \cite{MO} have used Monte
Carlo
simulations of the three-anyon problem which have enabled them to extract the
third
virial coefficient. In this way they have obtained the simple result
\beq
         A_3 = \frac{1}{36} + \frac{1}{12\pi^2}\sin^2\theta.
\label{A_3}
\eeq
Corrections are expected to go as $\sin^4\theta$, but are absent or so small
that they
cannot be determined with the present level of accuracy in the simulations.
Higher
virial coefficients will be very difficult to obtain by this method. $A_3$ is
plotted in
Fig.5b and is seen to take the common value $A_3 = 1/36$ derived previously
for fermions
and bosons.

The Monte-Carlo result is in agreement with analytic calculations
where the grand partition function has been obtained perturbatively to first
\cite{Ouvry1}
and second \cite{Ouvry2} order in the statistical angle. These results have
more recently
been confirmed in the Chern-Simons formulation of anyons \cite{AL} to the same
order in perturbation theory \cite{Valle}. Extracting the lowest virial
coefficients
and plotting the resulting pressure as function of temperature at fixed
density, one
finds as expected a curve in between the fermion and boson lines in Fig.1 as
long as
$\rho\Lambda^2 < 1$ \cite{SV}. It has been shown that the relation
${\cal E} = P$
between energy density and pressure is valid for anyons as it is for bosons and
fermions \cite{Tor}. One can thus also obtain the specific
heat for anyons at high temperatures \cite{SV}. Since the virial expansion
cannot
be used at low temperatures, one obtains no information about the
thermodynamics of anyons in this regime where quantum effects dominate.

Instead of relying on the virial expansion, one can try to use the pressure
as function
of fugacity in the discussion of the equation of state. To first order in
$\theta$,
this is found to be \cite{Ouvry1,Valle}
\beq
    \beta P = \frac{1}{\Lambda^2}\left[Li_2(1 - z) - {\theta\over\pi}\ln^2(1
- z)\right].
         \label{A.4}
\eeq
With the density $\rho = z\partial/\partial z(\beta P)$ we can plot the
pressure as
function of temperature with the fugacity as a parameter. The result is
shown in Fig.6
for two small values of the statistical angle. We see that when the
temperature gets
sufficiently small, the resulting pressure is no longer well-defined
\cite{Ouvry1}.
When $\theta$ becomes smaller, this happens at a corresponding lower
temperatures.
A similar breakdown of perturbation theory is also
seen using the analytical results \cite{Ouvry2,Valle} for the pressure valid
to second order in $\theta$. It can be traced back to a violation of
$d\rho/dz \ge 0$
which follows from general statistical mechanics for systems in thermodynamic
equilibrium \cite{Ouvry1}. This inequality is derived in the Appendix. Since
it is here
perturbatively violated, one can even get negative particle densities for
sufficiently
low temperatures. Higher order perturbative results will probably not give
much more
understanding of the physical phenomena causing this unexpected behaviour at
low
temperatures in the anyon gas.

When the statistical angle $\theta$ is small, anyons can be described as
bosons with a
hard-core repulsion. Such a system is known to become a superfluid at low
temperatures
\cite{Huang}. Similarly, anyons with $\theta \sim 2\pi$ behave as fermions
with a weak,
attractive potential. This gives rise to a pairing force between the fermions
so they
effectively become bosons which condense. If the particles are charged, we
then have a
superconducting system. These heuristic arguments are actually supported by
approximative and non-perturbative calculations of anyons at very low
temperatures
\cite{SC,FW2} and are probably the reason for the breakdown of the perturbative
equation of state in the same region.

\vspace{10mm}
\section{Discussion and conclusion }
Until the discovery of fractional statistics, the thermodynamics of ideal
quantum gases
in two dimensions was completely understood although a few detailed
properties have first
been unravelled during the last couple of years. But these gases are in
general
characterized by a statistical angle which describe anyons interpolating
between ordinary
bosons and fermions. With the recent calculations of the first few virial
coefficients
we have today a fairly good knowledge of the high-temperature behaviour of
the ideal
anyon gas. There are also many indications and arguments for the gas
undergoing a
transition into a superfluid phase at low temperatures. However, the details
of this
transition and the physical properties of the new phase are still to a
large extent
unknown.

Anyons have many very beautiful properties and there is every reason to
believe that in the coming years many of the present problems will be much
better
understood or even solved. These results for two-dimensional quantum gases
might even
some time in the future have practical consequences with the ever increasing
proliferation of new materials and structures in solid state physics.

We would like to to thank Professor S. Ouvry for several useful suggestions
and comments.

\vspace{15mm}

\section{Appendix}

For fermions at zero temperature the chemical potential is just the Fermi
energy
$\ve_F$. As in three dimensions it follows directly form the particle density
\be
    \rho = {m\over 2\pi\hbar^2}\int_0^{\ve_F}\!d\ve
\ee
which gives $\ve_F = 2\pi\rho\hbar^2/m$. The corresponding energy density is
then
\beq
     {\cal E}_0 = {m\over 2\pi\hbar^2}{1\over 2}\ve_F^2
                = {\hbar^2\over m}\pi\rho^2
\label{A.1}
\eeq
which also will be the pressure in the fermion gas at zero temperature.

At temperatures just above zero we can use the Sommerfeld expansion
\cite{Huang}. In the
calculation of the energy, we then find it to terminate after just two terms.
\beq
     {\cal E}_F=\frac{m}{2\pi\hbar^2}\left[\frac{1}{2}\mu^2+\frac{\pi^2}{6}
(kT)^2 \right]
                                                                    \label{A.2}
\eeq
Similarly, we find no power corrections to the chemical potential, i.e.
$\mu = \ve_F$ as long
as $k_BT \ll \ve_F$. This is also in agreement with the exact result
(\ref{2.7}) which gives
\beq
     \mu = \ve_F + k_B T\ln{\left(1 - e^{-\rho\Lambda^2}\right)}
\eeq
For the low-temperature energy density of the fermion gas we thus have
\beq
     {\cal E}_F = {\cal E}_0
     \left[1 +\frac{\pi^2}{3}\left(\frac{1}{\rho\Lambda^2}\right)^2 \right]
\label{A.3}
\eeq
where $\Lambda = (2\pi\beta\hbar^2/m)^{1/2}$ is the thermal wavelength. Since
we have
from (\ref{2.15}) that ${\cal E}_F - {\cal E}_B = {\cal E}_0$, we see that
the last term
in (\ref{A.3}) is the bosonic energy density ${\cal E}_B$ at low temperatures.

There are no corrections to these results involving higher orders in
$1/\rho\Lambda^2$
which is small at low temperatures. Instead, there will be exponential
corrections
of the order $\exp(-\rho\Lambda^2)$. One way to see this, is to use the
following
property of the dilogarithmic function \cite{AS}
\beq
     \mbox{Li}_2(x) + \mbox{Li}_2(1-x)= \frac{\pi^2}{6} - \ln x\ln(1-x)
\eeq
For the energy density (\ref{2.7a}) of the boson gas we then have
\beq
     {\cal E}_B & = & {1\over\beta\Lambda^2}\mbox{Li}_2\left(e^{-\rho
\Lambda^2}\right) \\
                & = & {{\cal E}_0\over(\rho\Lambda^2)^2}\left({\pi^2\over 3}
                  + 2\left[\rho\Lambda^2\ln{\left(1 -
e^{-\rho\Lambda^2}\right)}
                - \mbox{Li}_2\left(1 - e^{-\rho\Lambda^2}\right)\right]\right).
\eeq
This provides an exact and explicit result for the energy density of a
two-dimensional
gas of free bosons valid at all temperatures. Adding the constant
(\ref{A.1}), we then
get the corresponding exact result for the fermion gas.

We now want to show that for a gas in thermodynamic equilibrium the
derivative of the
density $\rho$ with respect to the to the fugacity $z = \exp{(\beta\mu)}$ is
always
positive. From the grand canonical partition function \cite{Huang}
\beq
     \Xi = \sum_{N=0}^\infty z^N Z_N
\eeq
where $Z_N$ is the canonical partition function for exactly $N$ particles,
we obtain the
average number of particles
\beq
     \ex{N} = {\del\ln{\Xi}\over\del\ln{z}}
            = {1\over\Xi}\sum_{N=0}^\infty N z^N Z_N.
\eeq
The density is $\rho = \ex{N}/V$ where $V$ is the volume of the system.
Taking  the derivative with repect to the fugacity at constant volume, yields
\be
      V{\del\rho\over\del z} = {1\over\Xi}\sum_{N=0}^\infty N^2 z^{N-1} Z_N
    - {1\over\Xi^2}\sum_{N=0}^\infty N z^N Z_N\sum_{N'=0}^\infty N' z^{N'-1}
Z_N'.
\ee
Combining the two terms, we get
\be
      {\del\rho\over\del z} = {1\over zV}\sum_{N,N'} z^{N+N'} Z_N Z_{N'}(N^2
- NN').
\ee
Because of the symmetry between $N$ and $N'$ we can rewrite this as
\be
      {\del\rho\over\del z} & = & {1\over zV}\sum_{N,N'} z^{N+N'} Z_N Z_{N'}
      (\2[N^2 + {N'}^2] - NN')  \\
                  & = & {1\over 2zV}\sum_{N,N'} z^{N+N'} Z_N Z_{N'}(N - N')^2
\ee
which is always a positive quantity.

To lowest order in perturbation theory we had the result (\ref{A.4}) for the
pressure in
the anyon gas as function of the fugacity. The corresponding density is
\beq
      \rho = {\del P\over\del\mu}
      = -\frac{1}{\Lambda^2}\ln(1 - z)\left[1 - {\theta\over\pi}{2z\over
1 - z}\right].
\eeq
The last term is seen to  give a violation of the above bound at low
temperatures where the fugacity approaches one.

\eject

{\bf Figure captions:}

Figure 1: Equations of state for free bosons and fermions in two dimensions
as functions of
$1/\rho\Lambda^2 \propto T/\rho$.

Figure 2: The exchange of two anyons generates a phase-factor which depends
on the direction of
the corresponding rotation in their center of mass.

Figure 3: Definition of the angle $\phi_{ij}$ in the positions of two anyons.

Figure 4: Spectrum of two anyons in a harmonic oscillator potential.

Figure 5: The second a) and third b) virial coefficient as a function of the
statistical
parameter $\alpha = \theta/\pi$.

Figure 6: The equation of state to first order in $\theta$ for $\theta$ = 0,
0.05, 0.1, with $\theta$ increasing from below.
The temperature at which the approximation breaks down, is seen to increase
with $\theta$.

\end{document}